\documentclass[apjl]{emulateapj}
\usepackage{amssymb,graphicx}
\usepackage{epsfig}
\usepackage{psfrag}
\usepackage[usenames]{color}
\usepackage{multirow}

\newcommand{\cqg}{Class.\ and Quant.\ Grav.}
\newcommand{\jcop}{J.\ Comp.\ Phys.}

\shorttitle{Eccentric BH-NS Mergers}
\shortauthors{Stephens, East, \& Pretorius}

\begin{document}

\title{Eccentric Black Hole-Neutron Star Mergers}

\author{Branson C.\ Stephens}
\affil{Center for Gravitation and Cosmology, University of Wisconsin-Milwaukee, Milwaukee, WI 53211, USA.}
\author{William E.\ East and Frans Pretorius}
\affil{Department of Physics, Princeton University, Princeton, NJ 08544, USA.}

\begin{abstract}
Within the next few years gravitational waves (GWs) from merging black holes 
(BHs) and neutron stars (NSs) may be directly detected, making a thorough 
theoretical understanding of these systems a high priority.  As an 
additional motivation, these systems may represent a subset of short-duration 
gamma-ray burst (sGRB) progenitors.  BH-NS mergers are 
expected to result from primordial, quasi-circular inspiral as well as
dynamically formed capture binaries.  The latter channel allows mergers with 
high eccentricity, resulting in a richer variety of outcomes.  We 
perform general relativistic simulations of BH-NS interactions with a range of impact parameters,
and find significant variation in the properties of these events that have
potentially observable consequences, namely the GW signature, remnant 
accretion disk mass, and amount of unbound material. 
\end{abstract}


\maketitle

\section{Introduction}
Merging binaries consisting of black holes (BHs) and
neutron stars (NSs) are prime targets for observation by 
ground-based gravitational wave (GW) detectors \citep[such as LIGO,][]{LIGO}
and may be the progenitors of some short-hard gamma-ray 
bursts \citep[sGRBs,][]{npp92}. 
The great diversity of sGRB characteristics 
and the potential variation in the corresponding GW signals motivates a 
thorough investigation of the possible outcomes of binary compact object (BCO) mergers.
BCOs may form through evolution of primordial binaries or through 
dynamical processes in star clusters~\citep{oleary,lee2010}.  The latter 
population motivates this study of
BH-NS interactions for systems which are initially marginally unbound.

Star clusters at the centers of galaxies undergo mass segregation, 
resulting in heavier objects concentrated toward the 
center \citep[see, e.g.][]{bw77}.  Recent Fokker-Plank models suggest 
that the Galactic nuclear cluster (NC) should have $\sim1800$ BHs and 
$\sim400$ NSs in the central 0.1~pc \citep{hopman06}. 
Such clusters are thus 
promising sites for BH-NS close encounters.  Using models
of galactic NCs, \citet{oleary} calculate the rate 
of binary BH formation through GW emission in close encounters.  
They find corresponding Advanced LIGO detection rates between 5 and 2700 per year,
and estimate that the BH-NS rate could be around 1\% of this.
These capture binaries form with relatively small periapsis separations, $r_p$; 
in particular, $\sim30$\% form with $r_p \lesssim 10 M$, where
$M$ is the total mass of the system (Fig.~4 of \citet{oleary}).  (Unless
otherwise stated we employ geometric units with $G=c=1$.)  This is due in
part to the large velocity dispersion in the cluster core 
($\sim1000$~km~s${}^{-1}$), but also to gravitational focusing, which 
may be understood as follows. The total rate 
is proportional to the cross-section:  $\Gamma \propto \pi b^2$, 
where $b$ is the impact parameter.  However, for Newtonian hyperbolic orbits 
with relative velocity $w$ at infinity, $r_p = b^2w^2/2M + O(w^4)$.  Thus, 
the rate is {\em linearly} proportional to $r_p$.

Globular clusters (GCs) that have undergone core collapse may also host 
BCO close encounters due to the high density of compact objects in their
cores \citep{fabian75,grindlay2006}.  For example, models of 
M15 calibrated to the observational velocity dispersion yield a 
NS fraction of $\sim55$\% in the inner 0.2~pc \citep{dull}.  
\citet{lee2010} calculate the expected rate of BCO interactions inside 
M15 as a function of time and then scale these results for GCs with a 
distribution of half-mass relaxation times.  
Depending upon the GC evolution model, they find that the global rate 
for BH-NS {\em collisions} (i.e., events for which $r_p \le R_{\rm NS}+R_{\rm BH}$) 
peaks at $\sim8-25~$yr${}^{-1}$Gpc${}^{-3}$
at redshifts between $z=0.36$ and $z=0.97$, and slowly declines
to between $50-85\%$ of peak by $z=0$.  (We obtained the BH-NS 
collision rate by re-scaling their NS-NS results according to the factors
in Table 3 of \citealt{lee2010}).  For the fiducial BH-NS system
considered by \citet{lee2010}, collisions occur at 
$r_p\leq2.7 M$.  Since the rate scales linearly with $r_p$, this implies
an interaction rate $\sim30-100$~yr${}^{-1}$Gpc${}^{-3}$ with 
(for example) $r_p\le10 M$.

Population synthesis models \citep{belczynski} find comparable
rates for {\em primordial} BH-NS mergers: from 
$\sim0.1~$yr~${}^{-1}$Gpc${}^{-3}$ (pessimistic) to 
$\sim120~$yr~${}^{-1}$Gpc${}^{-3}$ (optimistic).  However, primordial BH-NS 
binaries will enter the LIGO band with essentially zero eccentricity 
\citep{kowalska}.  Thus, GW signals from BH-NS close encounters 
should be readily distinguishable due to their significant eccentricities.  
We further note that the sGRB rate, 8-30~yr${}^{-1}$Gpc${}^{-3}$ 
\citep{guettapiran06}, is comparable to the primordial BH-NS rate, and somewhat less than
the expected NS-NS merger rate (30-400~yr${}^{-1}$Gpc${}^{-3}$ \citep{belczynski}).  
The estimates of \citet{lee2010} thus suggest that close encounters in clusters 
could contribute significantly to 
the sGRB progenitor population, especially if their emission is less tightly beamed than that of
primordial mergers \citep{grindlay2006}.  

BH-NS mergers in clusters with $r_p \lesssim 10 M$ will exhibit complicated 
behaviors probing the strong field regime of general relativity (GR). 
For stellar mass BH companions, one cannot treat the NS as a perturbation 
of the BH spacetime.
Furthermore, the non-linear nature of GR will most strongly
manifest during a close encounter of the BH-NS.
Numerical simulations within full GR are thus the preferred tool for exploring these systems. 
To date, such simulations have been performed by several
groups \citep{illinoisBHNSspin,shibataBHNS3,matt,chawla,pannarale} 
(see \citealt{mattsreview} for a review).  These studies employ
quasi-circular initial data, appropriate for primordial systems, and have 
explored a range of behaviors depending on the mass ratio, the BH 
spin, the NS equation of state (EOS), and the magnetic field.  
Our results complement these works
by offering a first study within full GR of initially
hyperbolic encounters, of relevance to BH-NS capture events and related
systems that merge with large eccentricity.

In the remainder of the letter we outline our numerical method, 
present results 
of our parameter space survey, and discuss some of their implications. 
The main result is the striking dependence of the outcome---disk mass,
unbound material, and GW signal---on the impact parameter. Though the
most ``extreme'' outcomes might require fine-tuning
and hence be rare, there is strong variation over the entire
range $r_p \lesssim 10M$ we considered: for a hyperbolic encounter
in a NC, this corresponds to roughly $30\%$ of encounters that lead
to a bound system. We were not able to follow the larger
$r_p$ cases through merger due to lack of computational resources, though
certainly a fraction of these should also exhibit similar variability
at the time of merger. This suggests that these systems could be 
a wellspring of varied and interesting GW and electromagnetic emission.

\section{Numerical approach}
Our 3D numerical code solves the Einstein field equations 
using finite difference (FD) techniques with Berger and Oliger style 
adaptive mesh refinement \citep[AMR,][]{bo84}.  The numerical scheme for evolving 
the spacetime metric is substantively the same as the generalized harmonic method
described by \citet{gh3d,paper2}, except that the FD scheme is fourth
order accurate and uses fourth order Runge-Kutta time integration.
We model the NS material as a perfect fluid and solve the 
hydrodynamic energy-momentum and continuity equations using conservative 
high-resolution shock-capturing schemes with second order Runge-Kutta time 
integration, and enforce strict conservation
at AMR boundaries using the flux correction method of \citet{bc89}.
We have implemented several 
methods for calculating inter-cell fluxes (HLL, \citealt{hll}; the Roe 
solver, \citealt{eulderink}; and the Marquina flux, \citealt{marquina}) and for 
reconstructing fluid primitive variables at cell interfaces (MC and 
minmod, \citealt{toro}; PPM, \citealt{ppm}; and WENO-5, \citealt{weno5}).
Unless otherwise noted, the simulations described below were performed with HLL and WENO-5.  
Our hydrodynamics scheme allows for any EOS of the form $P=P(\rho,\epsilon)$ (e.g., 
$\Gamma$-law, piecewise polytrope, and tabular EOSs).  
We have tested the new hydrodynamics sector of our code on problems 
including 1D and 2D Riemann problems, Bondi accretion, and single NSs.  
More details on our code and tests will be presented in \citet{upcomingpaper}.

For this first study the only parameter we vary is the initial periapsis separation.
The BH and NS have a 4:1 mass ratio, and both are
initially non-rotating.  The NS is modeled as a TOV 
star with a broken $\Gamma$-law EOS 
(labeled ``HB'' in \citealt{jocelyn} and including a thermal component to allow for shock heating) and has mass $1.35 M_\odot$ and radius 
11.6~km.  
The initial orbital parameters describe a hyperbolic encounter
with relative velocity $w=1000$~km~s${}^{-1}$, corresponding to the central 
region of a NC \citep{oleary}. These orbits are nearly parabolic, with 
$e-1\sim O(10^{-5})$, and hence the close-encounter behavior also
adequately describes such events in GCs.
We superimpose initial data for the BH and NS at an 
initial separation of $50 M$ (498~km) with initial velocities according to a
Newtonian orbit with the desired $r_p$.  
Though these superposed initial data do 
not strictly satisfy the constraint equations except at infinite separation, 
tests performed at various initial separations indicate that $50 M$ 
is sufficiently large that the constraint violation does not appreciably affect the 
system (relative to truncation error).

\section{Results and discussion}
We consider a range of periapsis separations
from $r_p/M=5.0$ to 15 ({\it i.e.,} 50 to 150 km). (Henceforth, we will 
consider $r_p$ to be normalized by $M$.)
In all of these cases, sufficient energy is carried 
away by GWs to result in a bound system.  Our simulations exhibit three types 
of behavior: (1) a direct plunge 
($r_p=5.0,5.83,6.67,6.81$), (2) following the initial periapsis
passage, a single elliptical orbit and then a plunge 
($r_p=6.95,7.22,7.5$), and (3) following the initial periapsis
passage, a long-period elliptical orbit ($r_p=8.75,10.0,12.5,15.0$).  For the
latter group (and the high resolution $r_p=7.5$ run), the entire orbit is prohibitively long to 
simulate, and we focus on the burst of GWs associated with the first periapsis passage.
For one case in each class ($r_p=5.0,7.5,10.0$) we ran three simulations
with different characteristic mesh spacings (but always with 7 refinement levels) for convergence studies.
At $t=0$, the low (medium, high) resolution 
run had finest meshes covering the BH and NS of roughly $80^3$ ($100^3$,$150^3$)
cells, resolving the NS diameter with $\sim40$ ($50$, $75$) cells
and the BH horizon diameter with $\sim70$ ($85$, $130$) cells.
(We note that the level structure is set by truncation error estimates and is adjusted with time.)

All other simulations were run at medium resolution. 
Unless otherwise noted, results will be reported for medium resolution, with 
error bars (where appropriate) computed from convergence
calculations.  
Our simulations employ compactified coordinates such that the 
outer boundaries extend to spatial infinity.  Thus, the global (ADM)
$M$ and $J$ should be conserved.  In practice, however, we must 
evaluate these quantities at a finite distance, making them
subject to gauge artifacts, some propagating outward from the central
BH/NS region from $t=0$. 
For $t<200M$, an extraction sphere of $300M$ is free of propagating artifacts, whence 
$M$ ($J$) is conserved to better than $0.3$ (2.0)\% 
for all cases at medium resolution. 

Based on the above of runs, some guidance from perturbative
results \citep{pm63,turner77,bg2010}, and a zoom-whirl geodesic analogue of the two
body problem \citep{Pretorius:2007jn}, we conjecture the following qualitative
behavior as a function of $r_p$ for initial encounters
resulting in a bound system. 

Consider $n$, the non-negative, integer number
of periapsis passages before disruption/plunge, and $r^i_p$, the
periapsis distance on the $i^{th}$ encounter for $1\le i \le n$. 
{\em Define} $r_p^{n+1}$ to be an {\em effective} periapsis distance for the final close encounter (i.e., the corresponding $r_p$ before the disruption/plunge part of the final encounter).
Group (1) above
has $n=0$, group (2) $n=1$, and group (3) $n\geq1$. 
The behavior of a close encounter will depend 
sensitively on the distance $\delta r^i_p=r^i_p-r_c$ between $r^i_p$ and a 
radius $r_c$ of an effective {\em unstable circular} orbit with the {\em same} 
energy and angular momentum. If $\delta r^i_p$
is sufficiently small (relative to $M$), the orbit will exhibit a whirl phase,
where it asymptotes to a nearly circular orbit.  
The smaller $\delta r^i_p$, the longer the duration of the whirl,
with a maximum when $\delta r^i_p=0$ that equals the time required for the binary
to lose its {\em excess} orbital kinetic energy, either via GW emission or tidal
transfer of energy to the NS material.
The excess energy is the difference between the total energy of the binary 
entering the whirl phase and a putative binary on a {\em quasi-circular} inspiral at $r_c$.

Because of the requirement that $\delta r^i_p$ be small, we only expect the possibility of significant whirling 
near the ultimate or penultimate encounter.
If $\delta r^i_p$ is negative, the whirl will directly transition to a plunge. If positive
and small, there will be a separation following the whirl; 
however the effective $r_c$ for the next encounter {\em increases} 
while $r^i_p$ {\em decreases} due to GW emission, and since this 
is quite sizeable for our 4:1 mass ratio
system, $\delta r^{i+1}_p$ will likely be negative, resulting in a subsequent plunge.
Furthermore, the separation where the NS starts to be tidally
disrupted is within the radii of unstable whirl orbits, hence the NS will not survive
any prolonged whirl phase. Prior encounters (for larger $n$ cases)
will not exhibit significant whirling, and while $\delta r^i_p$ is large 
the orbital evolution could better be described as a 
series of precessing ellipses with decreasing eccentricity and semi-major axis. 

Considering the number of orbits $n$ as a function of the {\em initial} $r_p$, 
$n(r_p)$ is monotonically increasing, and the values of $r_p$ where one could
see a notable final whirl would be near the steps in $n(r_p)$.
The most pronounced whirl behavior will occur for
small $n(r_p)$; as $n(r_p)$ increases, the amount of excess kinetic energy left over
once the final orbit is reached decreases, and for sufficiently large $n$ the
late stages will essentially follow a quasi-circular inspiral.

\begin{figure*}
\begin{center}
\includegraphics[height = 1.3 in]{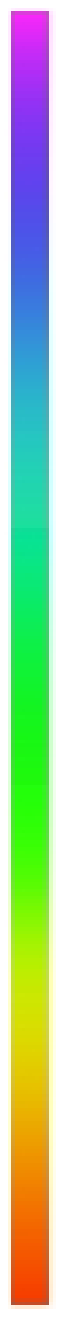} 
\put(2,90){$1\times \rho_{0}$} 
\put(2,45){$10^{-3}$}
\put(2,0){$10^{-6}$}
\hspace{0.5 in}
\includegraphics[clip=true, draft=false, viewport=0 0 780 665, width=1.66in]{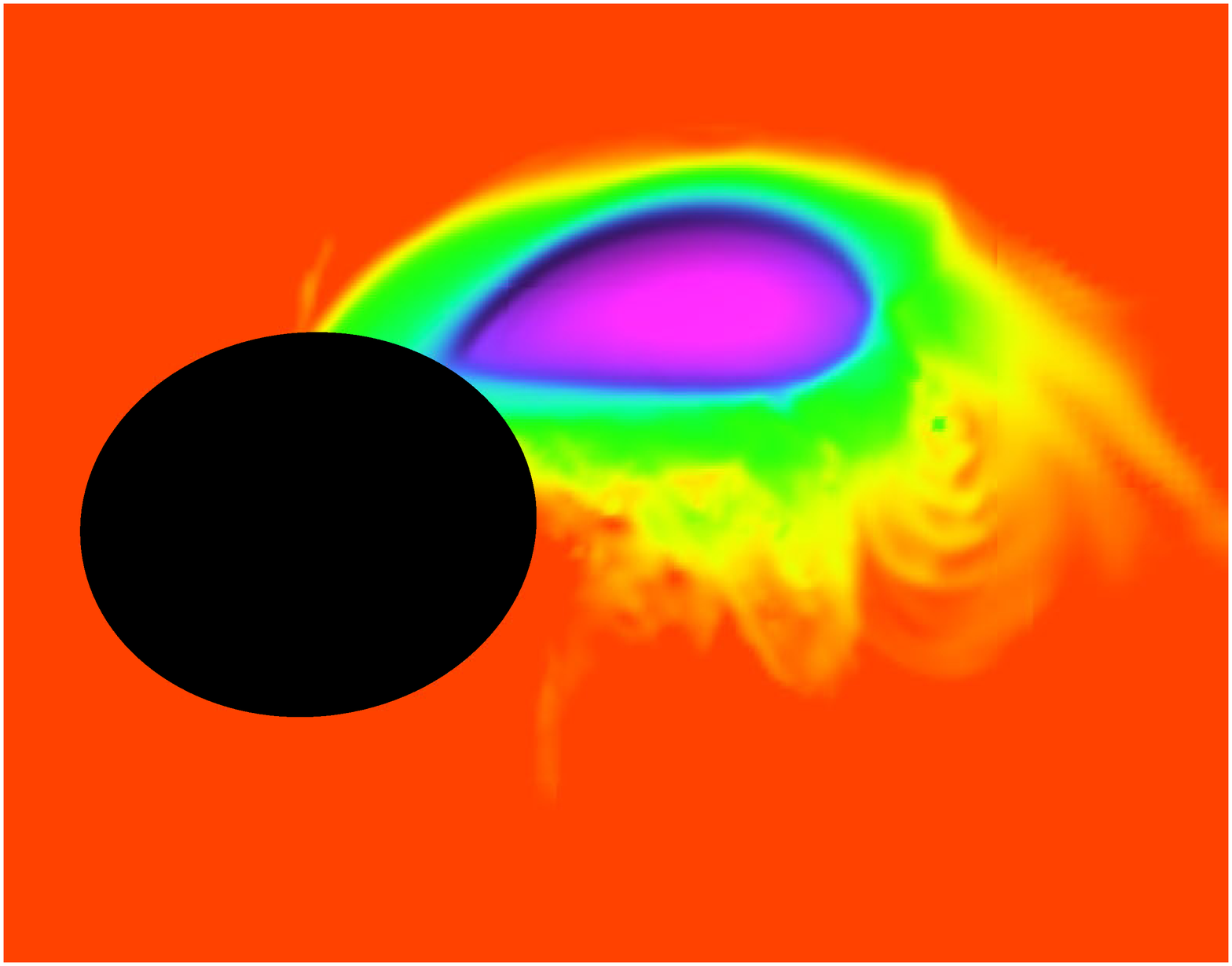} 
\includegraphics[clip=true, draft=false, viewport=0 0 780 615, width=1.66in]{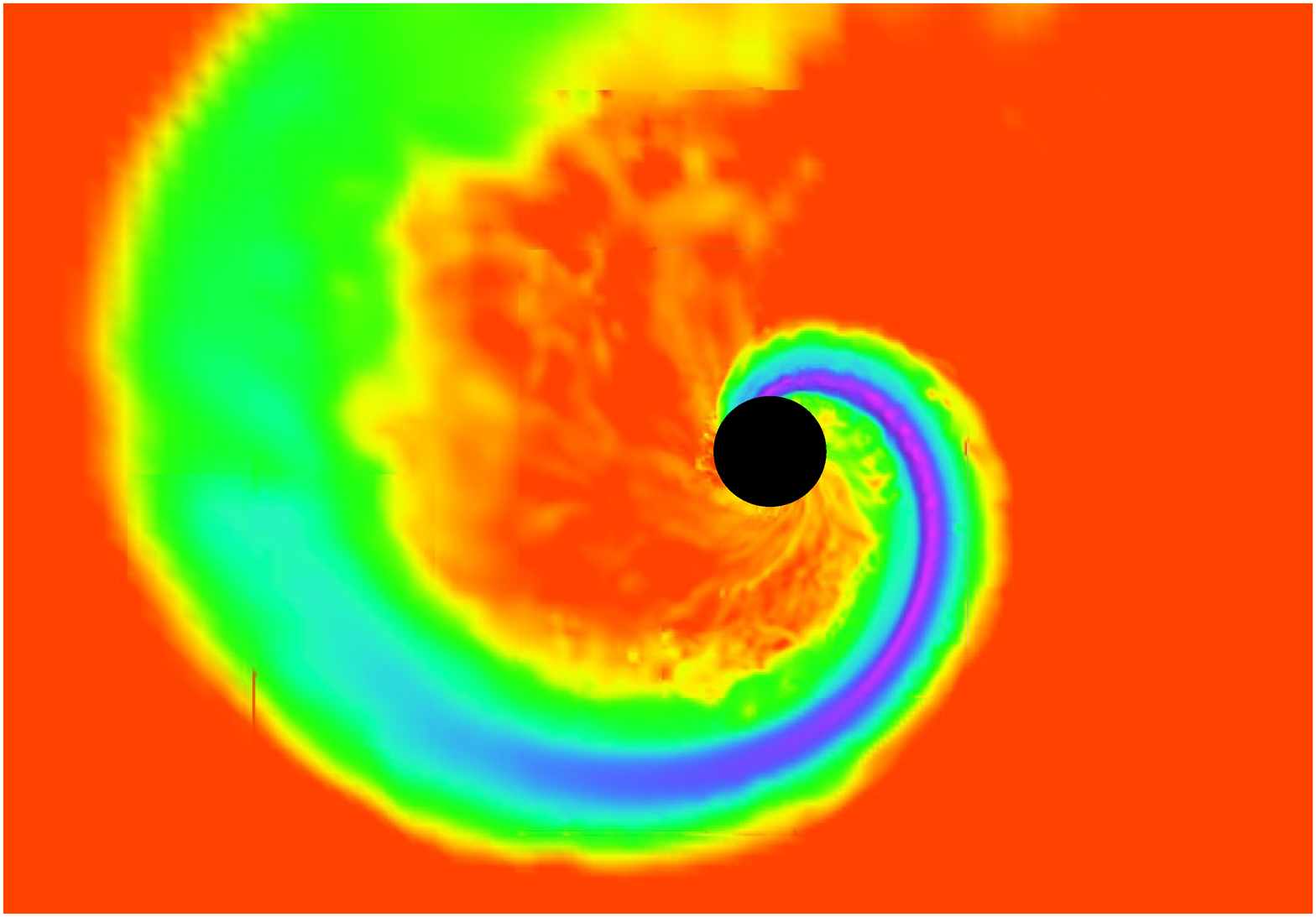} \hspace{0.4 in}
\includegraphics[height = 1.3 in]{vertical_scale.eps} 
\put(2,90){$10^{-4} \rho_{0}$} 
\put(2,45){$10^{-6}$}
\put(2,0){$10^{-8}$}
\hspace{0.5 in}
\includegraphics[clip=true, draft=false, viewport=20 0 832 640, width=1.66in]{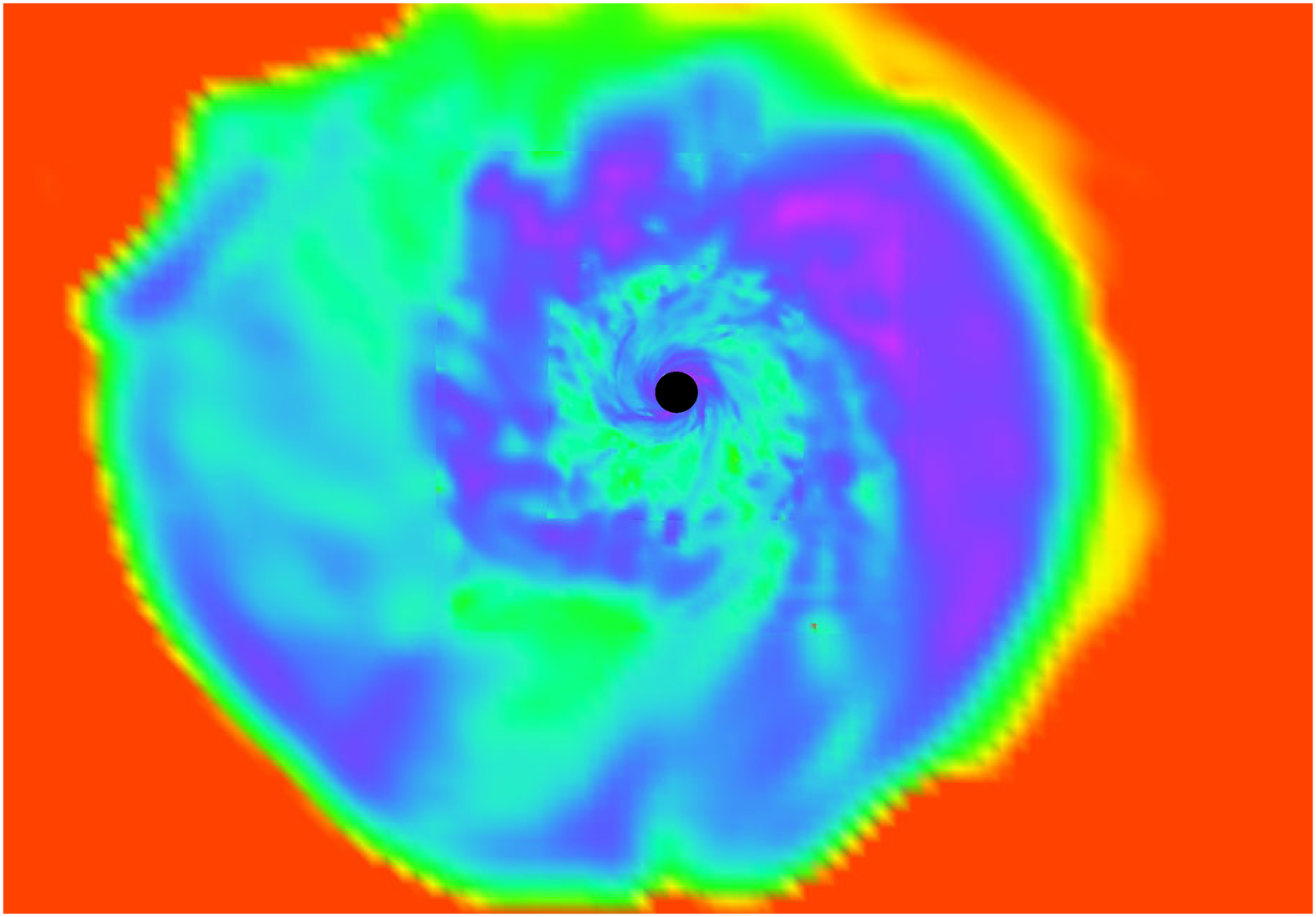} \\
\hspace{0.6 in}\includegraphics[clip=true, draft=false, viewport=30 30 810 645, width=1.66in]{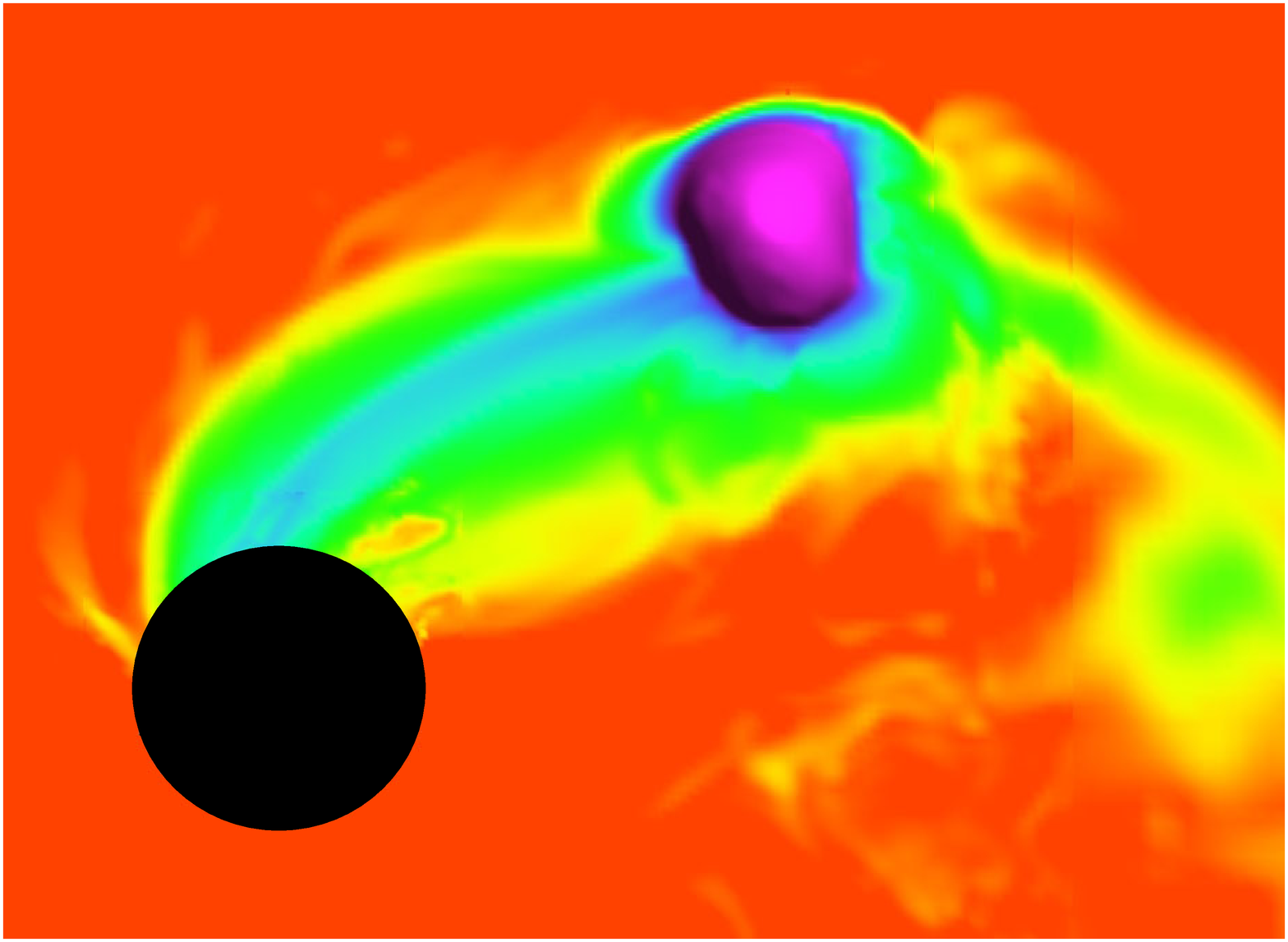}
\includegraphics[clip=true, draft=false, viewport=120 100 687.3 547.3, width=1.66in]{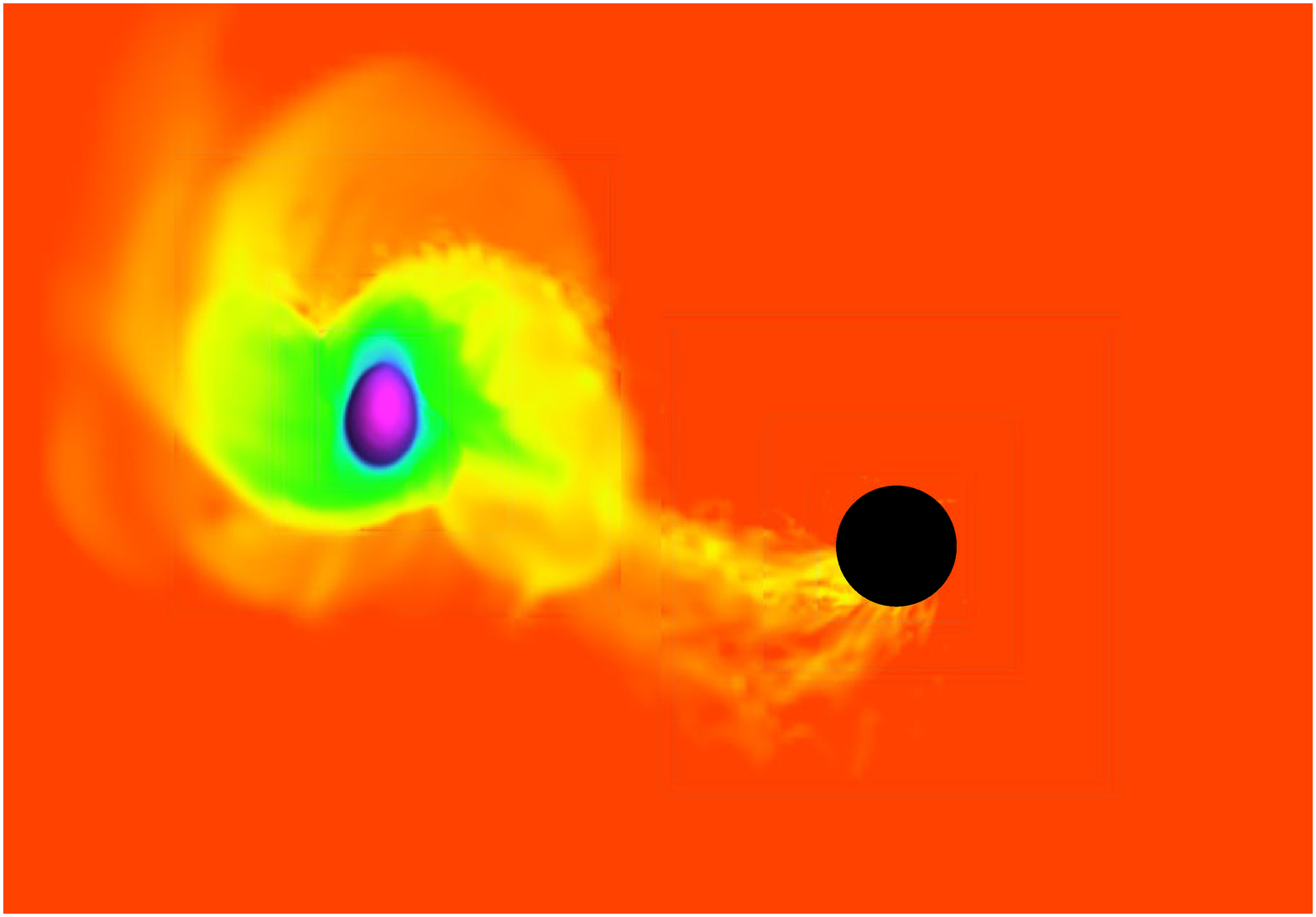} \hspace{0.4 in}
\includegraphics[height = 1.3 in]{vertical_scale.eps}
\put(2,90){$10^{-5} \rho_{0}$} 
\put(2,60){$10^{-6}$}
\put(2,30){$10^{-7}$}
\put(2,0){$10^{-8}$}
\hspace{0.5 in}
\includegraphics[clip=true, draft=false, viewport=0 0 780 615, width=1.66in]{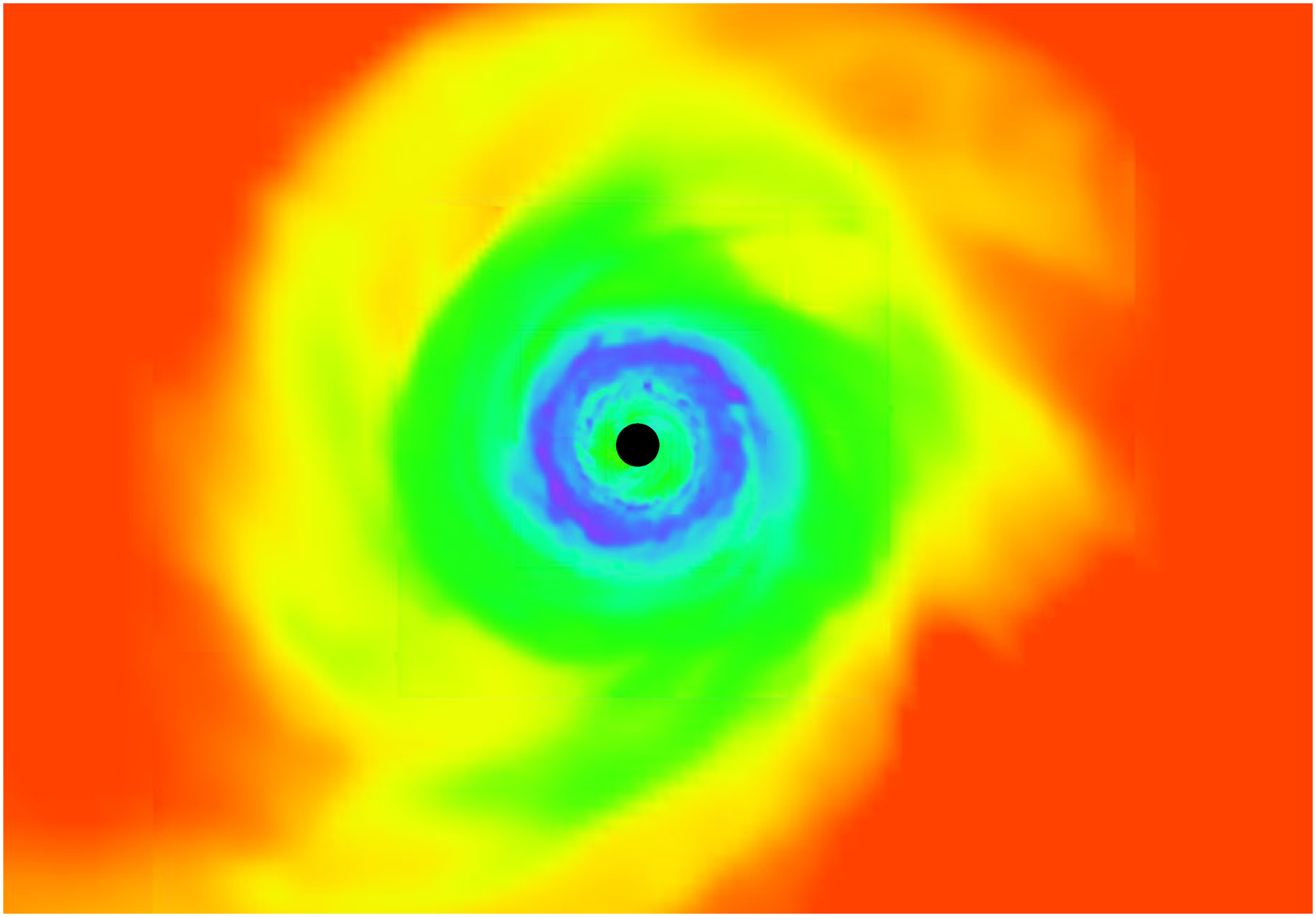}
\caption{Rest mass density in the equatorial plane from BH-NS simulations with varying $r_p$.  
The four panels on the left show (left to right, top to bottom, same color scale):   (1) the BH and NS merging ($t=6.82$~ms, $r_p=5$), 
(2) the NS being stretched into a long tidal tail ($t=9.97$~ms, $r_p=6.81$), 
(3) a brief mass transfer episode during the NS's first periapsis passage ($t=8.78$~ms, $r_p=6.95$), and
(4) the NS's subsequent distortion ($t=10.6$~ms, $r_p=6.95$).
On the right is 
a nascent accretion disk (top, $t=12.3$~ms,  $r_p=6.67$), and 
a late-stage accretion disk (bottom, $t=38.2$~ms, $r_p=7.5$, low resolution, PPM).  
The color scale is logarithmic with units of the initial maximum density ($\rho_{0}=8.3\times 10^{14}$ g cm$^{-3}$).
The black hole is roughly the same coordinate size in all panels ($R_{\rm BH}=$ 16 km), which can be used
to infer the relative scale of each snapshot.}
\label{snapshots}
\end{center}
\end{figure*}

Figure~\ref{snapshots} illustrates some of the varied phenomena encountered 
near the transition between $n(r_p)=0$ and $n(r_p)=1$, occurring
at $r_p\approx 6.88\pm0.08$. Most striking is
the amount of rest mass remaining after the merger as a function
of $r_p$ (Table~\ref{master_table}).  For cases in which the 
NS plunges without significant disruption, such as $r_p=5$ or $r_p=7.5$, 
less than 1\% of the initial mass is available to form an accretion
disk.  However, for $r_p$ closer to the first transition 
($r_p=6.67$ and 6.81, see Figure~\ref{snapshots}), 
the NS is stretched into a long tidal tail, and a sizeable
amount of bound material is left to form an accretion disk---12\% of 
the initial NS rest mass for $r_p=6.81$.

\begin{table*}
\begin{center}
{\small
\begin{tabular}{ l l l l l l l l }
\hline\hline
$r_p$      &  $M_0/M_0(t=0)$\tablenotemark{a}
 & $M_{0,u}/M_0(t=0)$\tablenotemark{b}
 & $\tau_{\rm acc}$ (ms)\tablenotemark{c}
 & \multicolumn{2}{c}{First periapsis\tablenotemark{d}}   
 & \multicolumn{2}{c}{Total\tablenotemark{e}}  \\
           &                &                   &                      & 
$\frac{E_{GW}}{M}\cdot10^2 $    &  $ \frac{J_{GW}}{M^2}\cdot10^2$ &  $ \frac{E_{GW}}{M}\cdot10^2$  & $\frac{J_{GW}}{M^2}\cdot10^2$  \\
\hline
5.00   &  0.005   &  0.0     &   25   &   \ldots         &  \ldots         &   $0.67(0.87)$ & $4.14(4.86)$ \\
6.67   &  0.107   &  0.056   &   130  &   \ldots         &  \ldots         &   $1.29$       & $9.10$ \\
6.81   &  0.221   &  0.101   &   40   &   \ldots         &  \ldots         &   $1.19$       & $9.60$ \\
6.95   &  0.018   &  0.003   &   47   &  $0.697$        &  $7.33$        &   $1.65$       & $13.9$\\
7.22   &  0.013   &  0.001   &   16   &  $0.358$        &  $4.48$        &   $1.18$       & $10.2$ \\
7.50   &  0.009   &  0.003   &   7.6  &  $0.242(0.147)$\tablenotemark{f} &  $3.44 (2.46)$ &   $1.03$       & $44.7$\\
8.75   &  \ldots  &  \ldots  & \ldots &  $0.073$        &  $1.58$        &   \ldots        & \ldots  \\
10.0   &  \ldots  &  \ldots  & \ldots &  $0.033(0.027)$ &  $0.97(0.88)$  &   \ldots        & \ldots  \\
12.5   &  \ldots  &  \ldots  & \ldots &  $0.011$        &  $0.46$        &   \ldots        & \ldots  \\
\hline
\end{tabular}
}
\caption{Disk properties and GW energy and angular momentum losses}
\tablenotetext{1}{Rest mass remaining outside the 
BH shortly after merger, normalized by the initial total rest mass.}
\tablenotetext{2}{Unbound rest mass estimated using local fluid velocities.} 
\tablenotetext{3}{Rough {\em initial} accretion timescale ($\tau_{\rm acc} = M_0/\dot{M_0}$) 
evaluated shortly after merger.}
\tablenotetext{4}{Energy and angular momentum lost to GWs during the first close 
encounter.}
\tablenotetext{5}{Total GW energy and angular momentum losses for cases which 
were followed through merger.}
\tablenotetext{6}{Results are from medium
resolution runs; values in parentheses are 
Richardson extrapolated estimates using low and high resolutions, where available.
Note that the relatively large error for $r_p=7.5$ (and to a lesser extent
$r_p=5,10$) is due in part 
to truncation error altering the actual periapsis by a small 
amount, and in this regime the GW emission is highly sensitive to 
binary separation (Figure~\ref{gw_fit}).}
\label{master_table} 
\end{center}
\end{table*}

\begin{figure}
\begin{center}
\psfrag{formula1}{$\dot{M} \ (M_{\odot}/s)$}
\psfrag{formula2}{$t^{-5/3}$}
\includegraphics[width=3.25in,clip=true]{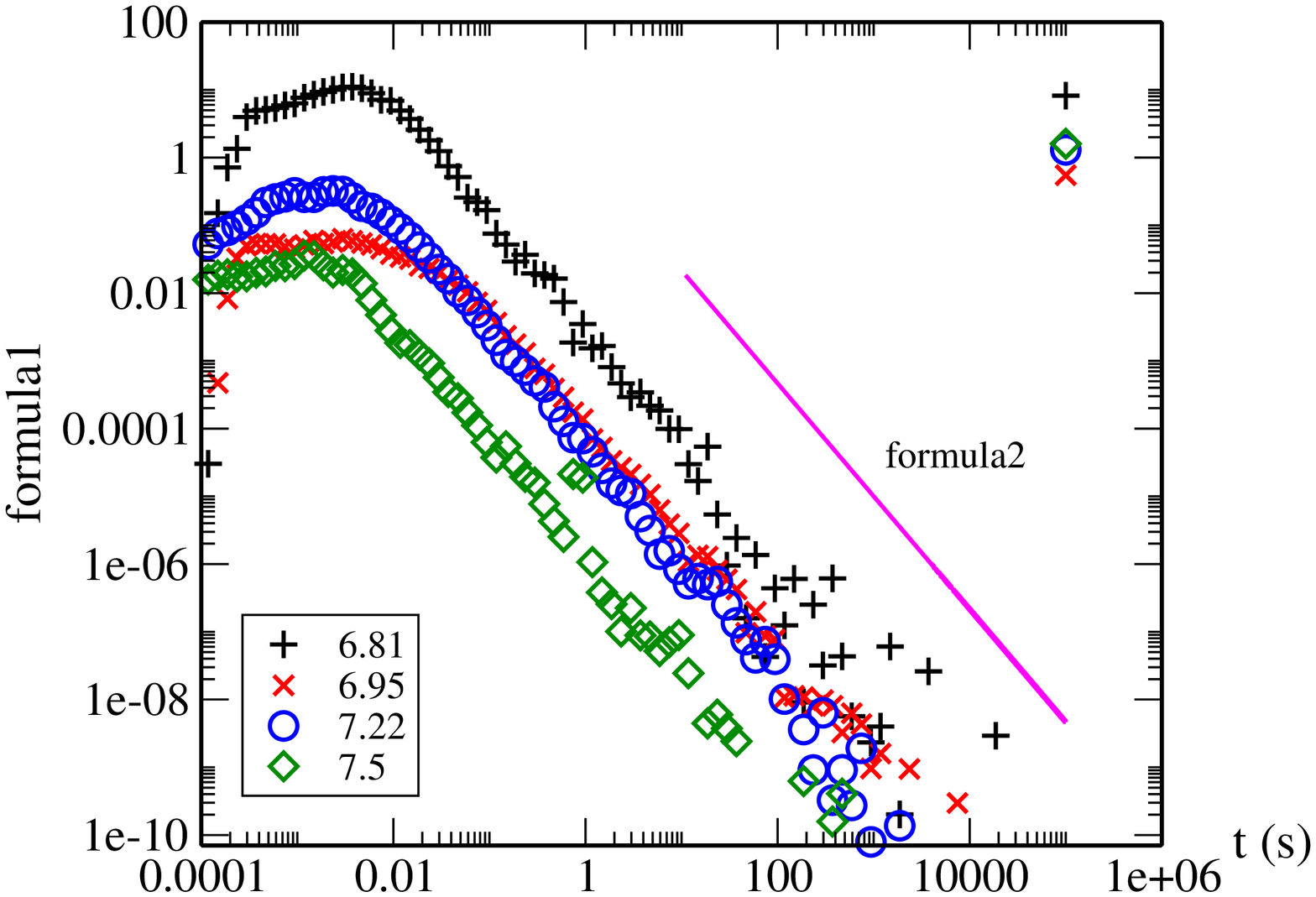}
\caption{Approximate fallback accretion rates for $r_p=6.81$, 6.95,
7.22, and 7.5.  These rates are evaluated at times ranging from 1.0 
to 2.2~ms after the approximate time when the BH accretion rate 
plateaus following the merger. 
For this diagnostic, we consider the fluid in each cell as a ballistic 
particle and take its orbital period as the approximate fallback
timescale.  The instantaneous BH accretion rates evaluated at the 
same time are shown at the upper right (arbitrary abscissa).} 
\label{fbplot}
\end{center}
\end{figure}

Figure~\ref{fbplot} shows the approximate rate of fallback 
as a function of time for $r_p=6.81$, 6.95, 7.22, and 7.5.  This is the 
rate at which material on elliptical orbits is expected to return to the
accretion disk \citep[see][]{rosswog07}.  (These accretion
rates are likely upper limits since they do not account for nuclear 
burning, see \citealt{metzger2}.) The fallback
rate is larger for cases with larger disk masses, such as $r_p=6.81$,
but all cases exhibit an approximate $t^{-5/3}$ falloff.  This time dependence
was predicted for stellar disruptions around supermassive BHs 
by \citet{rees88}.  It appears unlikely that BH-NS mergers with 
fallback rates as in Fig.~\ref{fbplot} will be able to explain sGRBs with 
extended emission  \citep[see, e.g.,][]{nb06} if this emission is due to feeding of the accretion
disk at late times.  For example, by $t\approx 100$~s, 
the luminosity for the $r_p=6.81$ case would be only 
$L\sim\eta\dot{M}c^2 \sim 2\times 10^{42}$~erg/s (assuming an efficiency 
$\eta=0.1$).
 
\begin{figure}
\begin{center}
\hspace{-0.5cm}
\includegraphics[height=1.5in,clip=true,draft=false]{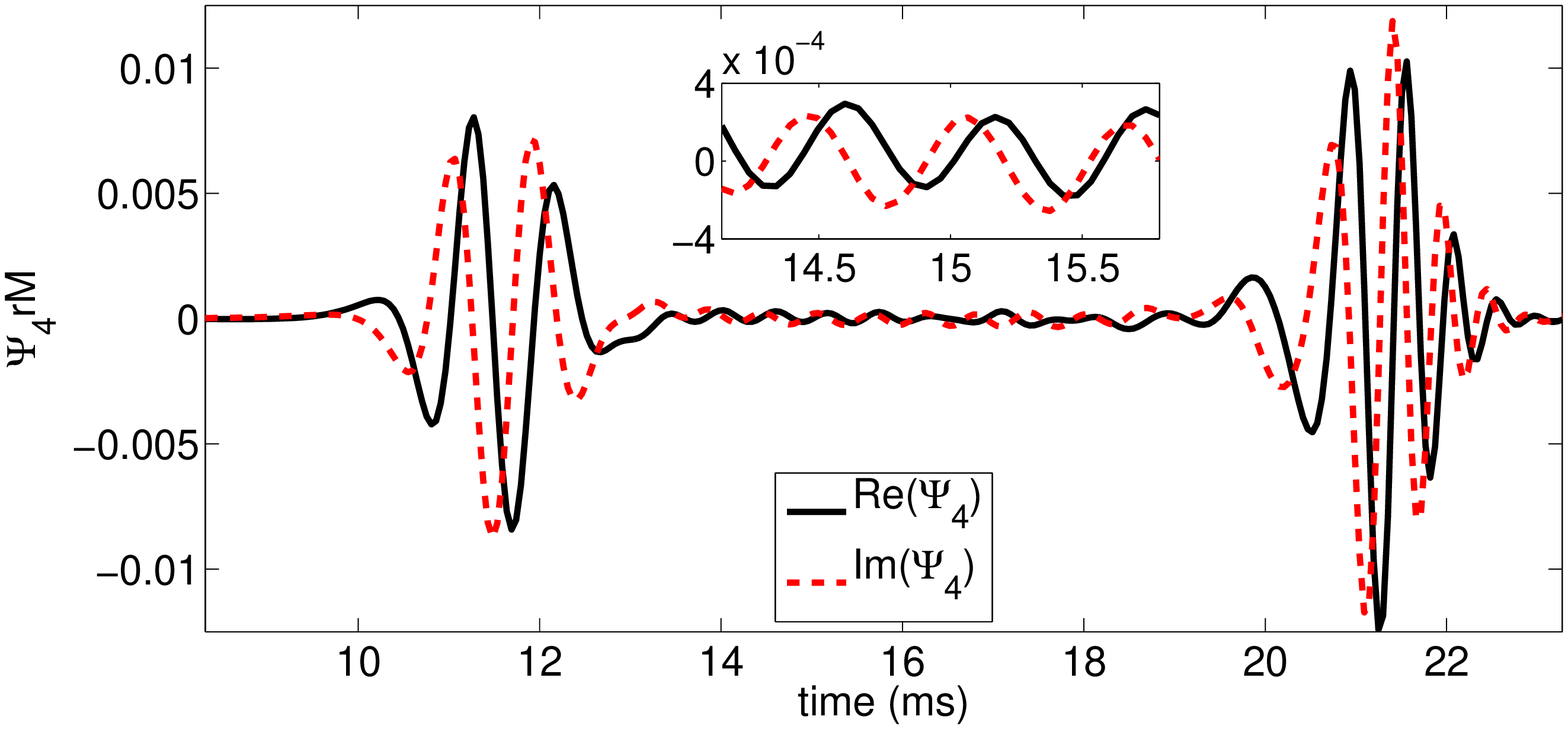}

\includegraphics[height=1.42in,clip=true,draft=false]{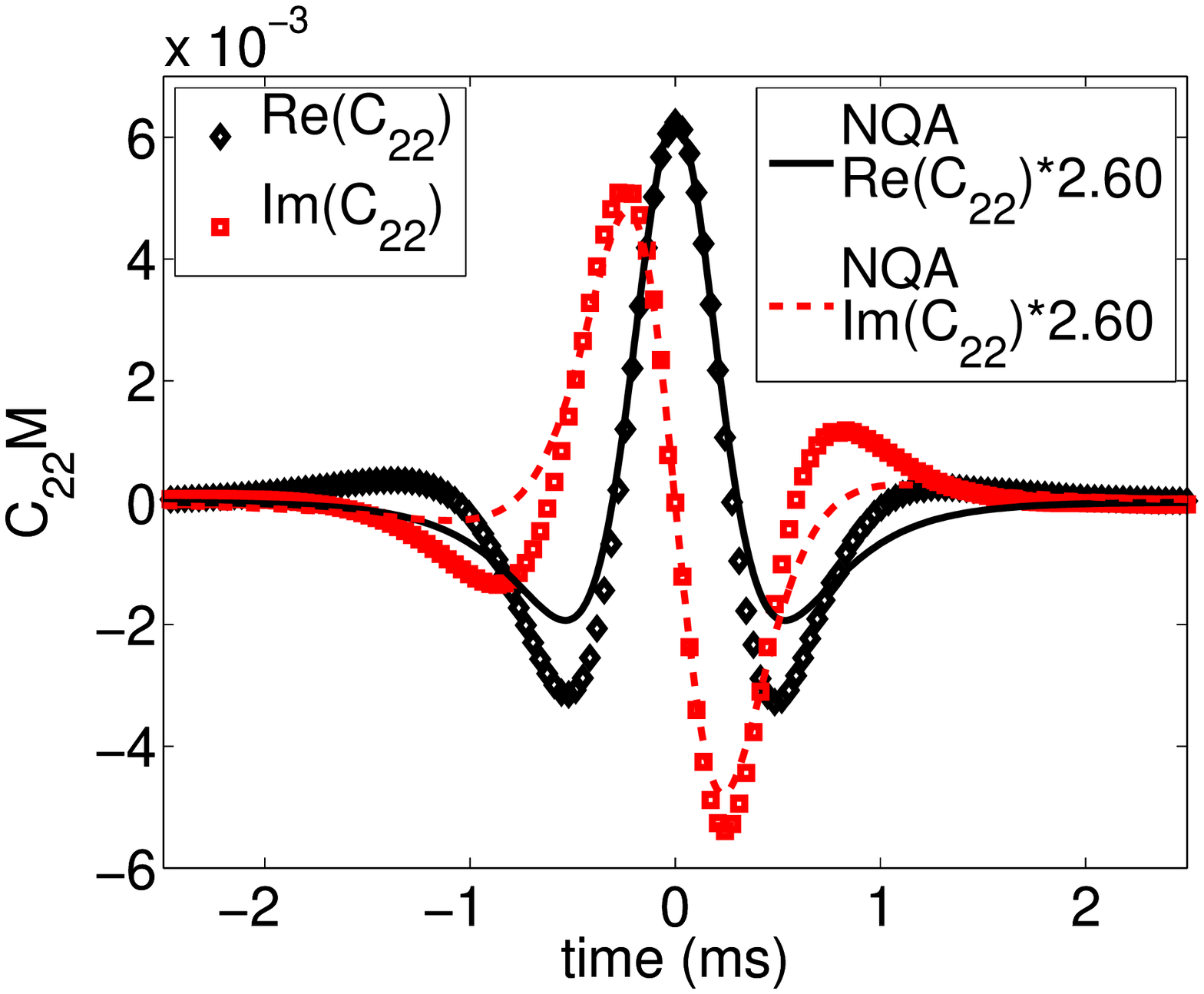}
\includegraphics[height=1.42in,clip=true,draft=false]{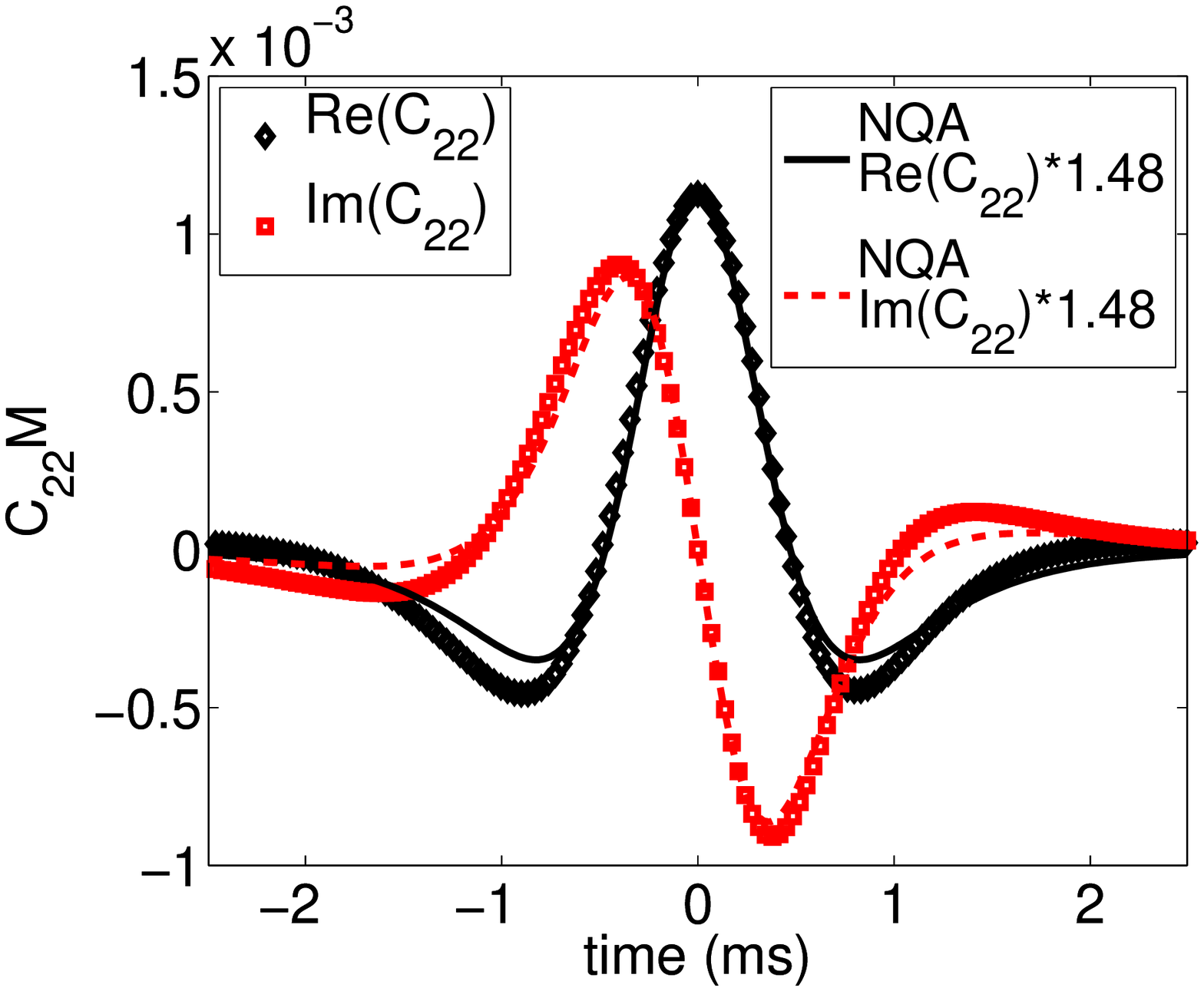}
\caption{ {\it (upper panel)}
The Newman-Penrose scalar $\Psi_4$ on the $z$-axis (orthogonal to the orbit) for $r_p=6.95$.  The first 
pulse is from the initial close encounter, the 
second from the merger-ringdown.  Between the pulses there is an oscillation due to 
the rotating, distorted neutron star, which is significantly torqued during the first encounter.  Here $t=0$
corresponds to the start of the simulation.
{\it (lower panels)} The real and imaginary components (black diamonds and red squares) of the $l=2$, $m=2$ 
spherical harmonic of $r\Psi_4$ for $r_p=7.5$ (left) and $r_p=10$ 
(right).  For comparison the NQA analytical 
results are shown multiplied by an overall factor so that the magnitude and 
phase match at peak ($t=0$).
}
\label{gwave_forms}
\end{center}
\end{figure}

\begin{figure}
\begin{center}
\includegraphics[height=2in,clip=true,draft=false]{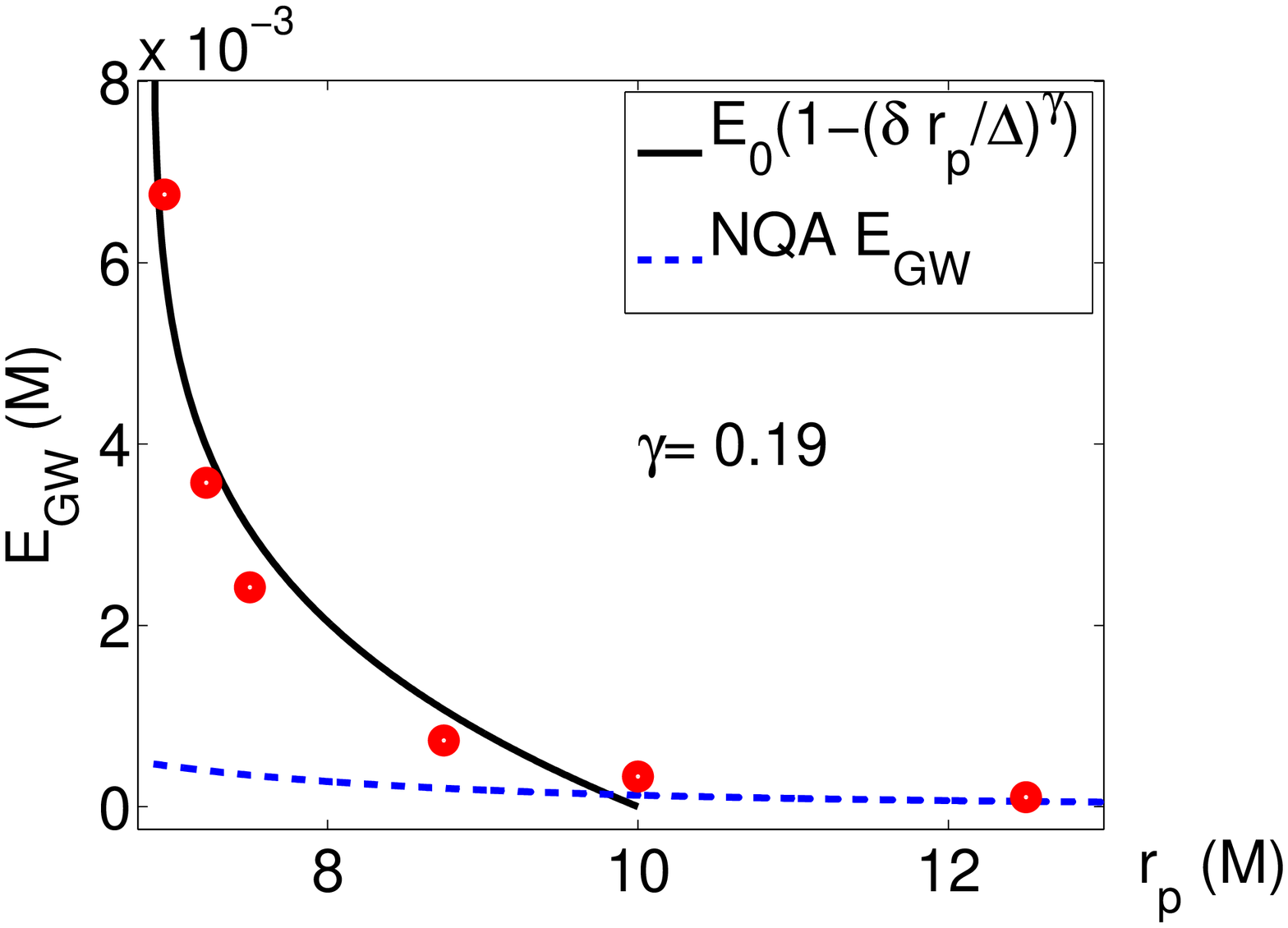}
\caption
{
Energy lost to GWs during the initial close encounter (i.e. excluding merger) as a function of $r_p$.  
The functional form $E_0(1-(\delta r_p/\Delta)^\gamma)$ (solid line), motivated
by zoom-whirl dynamics, is a fit
to the simulation results (red dots).
$\delta r_p=\delta r^1_p=r^1_p-r_c$ as discussed in the text; here $r^1_p=r_p$.
$E_0$ is the difference in energy between a 
quasi-circular orbit and an $e\approx 1$ orbit both with 
$r_p=r_c$.  $\Delta$ is the range over which zoom-whirl like behavior
dominates the GW emission energetics. $\gamma$ is a parameter
that in the geodesic analogue is related to the
instability exponent of the corresponding unstable circular
orbit; here, we use it as our fitting parameter. 
The NQA approximation is the dotted line.
}
\label{gw_fit}
\end{center}
\end{figure}

Table~\ref{master_table} also shows the total energy and angular momentum lost to 
GWs for $r_p$ between 5 and 12.5.  
For the cases that we followed through merger, we find
0.7-1.7\% of the total mass lost to GW energy, and estimate
the final spins of the BHs to be 
$(0.49\pm 0.01,0.45,0.37,0.47,0.50,0.50)$ for $r_p=(5.00,6.67,6.81,6.95,7.22,7.50)$ respectively.
The energy loss is largest
for the transitional case ($r_p=6.95$), which has a large pulse from 
the whirling first passage and a second burst from the merger (Fig.~\ref{gwave_forms}).  
Table~\ref{master_table} also shows the GW losses for the initial encounter in 
cases where the NS survives the periapsis passage (columns 5 and 6).  
These fly-by pulses 
can be compared with the prediction of \citet{turner77} 
(Fig.~\ref{gwave_forms}, lower panels, and Fig.~\ref{gw_fit}), who used the Newtonian
orbit together with quadrupole physics for the GW 
emission, which we will call the {\em Newtonian Quadrupole Approximation} (NQA).
Our waveforms show roughly the 
same pulse shape as the NQA prediction but have larger amplitudes for the smaller
$r_p$ cases.  At $r_p=15$ we find the gravitational waveform from the initial fly-by to be indistinguishable 
from the NQA prediction at our resolution ($\pm 10$\%).

The enhancement in GW energy losses for close encounters 
may be due (in part) to zoom-whirl-like behavior. 
Figure~\ref{gw_fit} shows the GW energy loss as a function 
of $r_p$, along with the NQA prediction and
a fit consistent with zoom-whirl dynamics \citep{Pretorius:2007jn} 
in the regime ($r_p \lesssim 10$) where we start to see significant departures
from the NQA approximation. 

\section{Conclusions}
An interesting result of this general relativistic study that 
is qualitatively consistent with previous Newtonian studies \citep[e.g.][]{lee2010}, 
is the great variability of the outcome as a function of impact parameter. 
For example, the remnant disk masses following merger range from
nearly zero up to $\sim0.3 M_{\odot}$.  
Mergers leading to significant disks occur in a small (but not negligible)
region of parameter space and could produce a sGRB.

In follow up work we plan to extend the parameter space survey,
varying the NS EOS and BH spin. We also intend
to explore the detectability of these events with GW 
detectors, and conclude with brief preliminary comments. These 
signals may be difficult to detect with 
instruments such as LIGO since they lack a long inspiral phase and most 
of the power is at high frequencies (1500-2000 Hz for the masses considered 
here).  Using the broadband AdLIGO noise 
curve,
we find sky-averaged SNR of 3-9 for $r_p=5-10$, assuming a distance of 100~Mpc.  
Scaling the GW emission up to $(M_{\rm NS},M_{\rm BH})=(2,8)M_{\odot}$ and 
assuming optimal orientation gives SNR of 8 out to 340~Mpc for $r_p=7.5$.  
Encounters with larger $r_p$ could be easier to detect since they will
have a number of fly-by GW bursts before the merger.  
Given that some BH-NS capture
binaries emit the majority of their GW power at the high frequency end
of the LIGO noise curve and 
could be relatively weak (both in terms of GW emission and extrapolated
electromagnetic emission based on disk mass), it may be a worthwhile
exercise to revisit the analysis of GRB 070201 \citep{Abbott:2007rh}
with burst-templates adapted to capture driven BH-NS encounters.

\acknowledgments
We thank Adam Burrows, John Friedmann, Roman Gold, Benjamin Lackey, and 
Richard O'Shaughnessy for useful conversations.  
This research was supported by the NSF
through TeraGrid resources provided by NICS under grant 
TG-PHY100053, the Bradley Program fellowship (BCS), the NSF 
Graduate Research Fellowship under grant DGE-0646086 (WE), NSF grants 
PHY-0745779 (FP) and PHY-1001515 (BCS),
and the Alfred P. Sloan Foundation (FP). 
Simulations were also run on the {\bf Woodhen} cluster 
at Princeton University.


\begin{thebibliography}{40}
\expandafter\ifx\csname natexlab\endcsname\relax\def\natexlab#1{#1}\fi

\bibitem[{Abbott {et~al.}(2008)}]{Abbott:2007rh}
Abbott, B., {et~al.} 2008, Astrophys. J., 681, 1419

\bibitem[{Abramovici {et~al.}(1992)}]{LIGO}
Abramovici, A., {et~al.} 1992, Science, 256, 325

\bibitem[{{Bahcall} \& {Wolf}(1977)}]{bw77}
{Bahcall}, J.~N., \& {Wolf}, R.~A. 1977, \apj, 216, 883

\bibitem[{{Belczynski} {et~al.}(2010){Belczynski}, {Dominik}, {Bulik},
  {O'Shaughnessy}, {Fryer}, \& {Holz}}]{belczynski}
{Belczynski}, K., {Dominik}, M., {Bulik}, T.,  {et~al.} 2010, \apjl, 715, L138

\bibitem[{Berger \& Colella(1989)}]{bc89}
Berger, M., \& Colella, P. 1989, \jcop, 82, 64

\bibitem[{Berger \& Oliger(1984)}]{bo84}
Berger, M.~J., \& Oliger, J. 1984, \jcop, 53, 484

\bibitem[{Berry \& Gair(2010)}]{bg2010}
Berry, C.P.L., \& Gair, J. 2010, \prd, 82, 107501

\bibitem[{{Chawla} {et~al.}(2010){Chawla}, {Anderson}, {Besselman}, {Lehner},
  {Liebling}, {Motl}, \& {Neilsen}}]{chawla}
{Chawla}, S., {Anderson}, M., {Besselman}, M., {et~al.} 2010, \prl, 105, 111101

\bibitem[{{Colella} \& {Woodward}(1984)}]{ppm}
{Colella}, P., \& {Woodward}, P.~R. 1984, \jcop, 54,
  174

\bibitem[{{Duez}(2010)}]{mattsreview}
{Duez}, M.~D. 2010, \cqg, 27, 114002

\bibitem[{{Duez} {et~al.}(2008){Duez}, {Foucart}, {Kidder}, {Pfeiffer},
  {Scheel}, \& {Teukolsky}}]{matt}
{Duez}, M.~D., {Foucart}, F., {Kidder}, L.~E., {et~al.} 2008, \prd, 78, 104015

\bibitem[{{Dull} {et~al.}(1997){Dull}, {Cohn}, {Lugger}, {Murphy}, {Seitzer},
  {Callanan}, {Rutten}, \& {Charles}}]{dull}
{Dull}, J.~D., {Cohn}, H.~N., {Lugger}, P.~M., {et~al.} 1997, \apj,
  481, 267

\bibitem[{{Etienne} {et~al.}(2009){Etienne}, {Liu}, {Shapiro}, \&
  {Baumgarte}}]{illinoisBHNSspin}
{Etienne}, Z.~B., {Liu}, Y.~T., {Shapiro}, S.~L., \& {Baumgarte}, T.~W. 2009,
  \prd, 79, 044024

\bibitem[{{Eulderink} \& {Mellema}(1995)}]{eulderink}
{Eulderink}, F., \& {Mellema}, G. 1995, \aaps, 110, 587

\bibitem[{{Fabian} {et~al.}(1975){Fabian}, {Pringle}, \& {Rees}}]{fabian75}
{Fabian}, A.~C., {Pringle}, J.~E., \& {Rees}, M.~J. 1975, \mnras, 172, 15

\bibitem[{{Grindlay} {et~al.}(2006){Grindlay}, {Portegies Zwart}, \&
  {McMillan}}]{grindlay2006}
{Grindlay}, J., {Portegies Zwart}, S., \& {McMillan}, S. 2006, Nature Physics,
  2, 116

\bibitem[{{Guetta} \& {Piran}(2006)}]{guettapiran06}
{Guetta}, D., \& {Piran}, T. 2006, \aap, 453, 823

\bibitem[{{Harten} {et~al.}(1983){Harten}, {Lax}, \& {van Leer}}]{hll}
{Harten}, A., {Lax}, P., \& {van Leer}, B. 1983, SIAM Rev., 25, 35

\bibitem[{{Hopman} \& {Alexander}(2006)}]{hopman06}
{Hopman}, C., \& {Alexander}, T. 2006, \apjl, 645, L133

\bibitem[{{Kowalska} {et~al.}(2011){Kowalska}, {Bulik}, {Belczynski},
  {Dominik}, \& {Gondek-Rosinska}}]{kowalska}
{Kowalska}, I., {Bulik}, T., {Belczynski}, K., {Dominik}, M., \&
  {Gondek-Rosinska}, D. 2011, \aap, 527, A70

\bibitem[{{Kyutoku} {et~al.}(2010){Kyutoku}, {Shibata}, \&
  {Taniguchi}}]{shibataBHNS3}
{Kyutoku}, K., {Shibata}, M., \& {Taniguchi}, K. 2010, \prd, 82, 044049

\bibitem[{{Lee} {et~al.}(2010){Lee}, {Ramirez-Ruiz}, \& {van de Ven}}]{lee2010}
{Lee}, W.~H., {Ramirez-Ruiz}, E., \& {van de Ven}, G. 2010, \apj, 720, 953

\bibitem[{{Marquina} {et~al.}(1992){Marquina}, {Marti}, {Ibanez}, {Miralles},
  \& {Donat}}]{marquina}
{Marquina}, A., {Marti}, J.~M., {Ibanez}, J.~M., {Miralles}, J.~A., \& {Donat},
  R. 1992, \aap, 258, 566

\bibitem[{{Metzger} {et~al.}(2010){Metzger}, {Arcones}, {Quataert}, \&
  {Mart{\'{\i}}nez-Pinedo}}]{metzger2}
{Metzger}, B.~D., {Arcones}, A., {Quataert}, E., \& {Mart{\'{\i}}nez-Pinedo},
  G. 2010, \mnras, 402, 2771

\bibitem[{Narayan {et~al.}(1992)Narayan, Paczynski, \& Piran}]{npp92}
Narayan, R., Paczynski, B., \& Piran, T. 1992, \apj, 395, L83

\bibitem[{{Norris} \& {Bonnell}(2006)}]{nb06}
{Norris}, J.~P., \& {Bonnell}, J.~T. 2006, \apj, 643, 266

\bibitem[{{O'Leary} {et~al.}(2009){O'Leary}, {Kocsis}, \& {Loeb}}]{oleary}
{O'Leary}, R.~M., {Kocsis}, B., \& {Loeb}, A. 2009, \mnras, 395, 2127

\bibitem[{{Pannarale} {et~al.}(2011){Pannarale}, {Tonita}, \&
  {Rezzolla}}]{pannarale}
{Pannarale}, F., {Tonita}, A., \& {Rezzolla}, L. 2011, \apj, 727, 95

\bibitem[{{Peters} \& {Mathews}(1963)}]{pm63}
{Peters}, P., \& {Mathews}, J. 1963, Phys.\ Rev., 131, 435

\bibitem[{{Pretorius}(2005{\natexlab{a}})}]{paper2}
{Pretorius}, F. 2005{\natexlab{a}}, \prl, 95, 121101

\bibitem[{{Pretorius}(2005{\natexlab{b}})}]{gh3d}
---. 2005{\natexlab{b}}, \cqg, 22, 425

\bibitem[{Pretorius \& Khurana(2007)}]{Pretorius:2007jn}
Pretorius, F., \& Khurana, D. 2007, \cqg, 24, S83

\bibitem[{{Read} {et~al.}(2009){Read}, {Markakis}, {Shibata}, {Ury{\= u}},
  {Creighton}, \& {Friedman}}]{jocelyn}
{Read}, J.~S., {Markakis}, C., {Shibata}, M., {et~al.} 2009, \prd, 79, 124033

\bibitem[{{Rees}(1988)}]{rees88}
{Rees}, M.~J. 1988, \nat, 333, 523

\bibitem[{{Rosswog}(2007)}]{rosswog07}
{Rosswog}, S. 2007, \mnras, 376, L48

\bibitem[{Stephens {et~al.}(2011)Stephens, East, \& Pretorius}]{upcomingpaper}
Stephens, B.~C., East, W.~E., \& Pretorius, F. 2011, in prep.

\bibitem[{{Tchekhovskoy} {et~al.}(2007){Tchekhovskoy}, {McKinney}, \&
  {Narayan}}]{weno5}
{Tchekhovskoy}, A., {McKinney}, J.~C., \& {Narayan}, R. 2007, \mnras, 379, 469

\bibitem[{Toro(1997)}]{toro}
Toro, E. 1997, Riemann Solvers and Numerical Methods for Fluid Dynamics
  (Berlin, Germany: Springer)

\bibitem[{{Turner}(1977)}]{turner77}
{Turner}, M. 1977, \apj, 216, 610

\end{thebibliography}
\end{document}